\def\funp{{I\!\!P}}
\def\xp{x_{{I\!\!P}}}
\newcommand{\beeq}{\begin{eqnarray}}
\newcommand{\eeeq}{\end{eqnarray}}
\def\ovbeta{{\beta}^\prime}
\documentclass{ws-p8-50x6-00}

\begin{document}

\title{DIFFRACTIVE PARTON DISTRIBUTIONS AND THE SATURATION MODEL}

\author{K. GOLEC--BIERNAT}

\address{II Institute of Theoretical Physics, Hamburg University,\\
Luruper Chaussee 149, D-22761 Hamburg, Germany\\ {\rm and}\\
Institute of Nuclear Physics, Radzikowskiego 152, \\
31-342 Krak\'ow, Poland}

\author{ M. W\"USTHOFF}

\address{ Department of Physics, University of Durham,\\ Durham DH1 3LE,
   United Kingdom}  


\maketitle

\abstracts{We construct diffractive parton distributions in the model
of DIS diffraction in which the diffractive state is formed by the
$q\bar{q}$ and $q\bar{q}g$ components. The interaction of such systems with
the proton is described by the dipole cross section given by the 
saturation model.
We find Regge factorization property with the measured at HERA depenedence
on energy. The found parton distributions are evolved with the DGLAP evolution
equations to make a comparison with the data.
}

\section{Introduction}

Collinear factorization proved for diffractive DIS~\cite{COL98} allows to
define diffractive quark and gluon distributions through the relation
to the diffractive structure function, e.g. in the leading $\log(Q^2)$
approximation
\be
\label{eq:dpd2}
F_2^{D(3)}\,=\,2\,
\sum_{q} e_q^2\;\beta\; q^D(\beta,Q^2,\xp),
\ee
where $q^D={\bar{q}}^D$. This relation concerns only the leading
twist description. The diffractive parton distributions (DPD)
obey the DGALP evolution equations. In the Ingelman-Schlein approach to DIS
diffraction {\it Regge factorization} is additionally assumed
\be
q^D(\beta,Q^2,\xp)\,=\,f(\xp)\,f_q(\beta,Q^2),
\ee
where $f(\xp)$ is related to the soft pomeron flux~\cite{H197,ZEUS99}.
In an alternative approach, the detailed description of diffractive processes
is achieved by modelling the  diffractive
final state as well as its interaction with the proton
starting from perturbative QCD~\cite{REV}. Such an analysis 
goes beyond the leading twist description and Regge factorization for the DPD 
is not assumed. The problem which is faced in this approach is the strong 
sensitivity to 
nonperturbative effects due to the dominance of the aligned jet configurations.
Thus, a model of soft or semihard interactions between the diffractive system 
and the proton is necessary. 

In our analysis we use 
the saturation model~\cite{GBW1} for this purpose which turned out to be 
successful in the description of inclusive and diffractive DIS data.
Extracting the DPD we are able to make a contact with the Ingelman-Schlein 
approach. We  find Regge factorization with the measured energy dependence and 
quantify the role of the higher twist contribution.

\section{Diffractive parton distributions}

\begin{figure}[t]
  \vspace*{-0.5cm}
     \centerline{
         \epsfig{figure=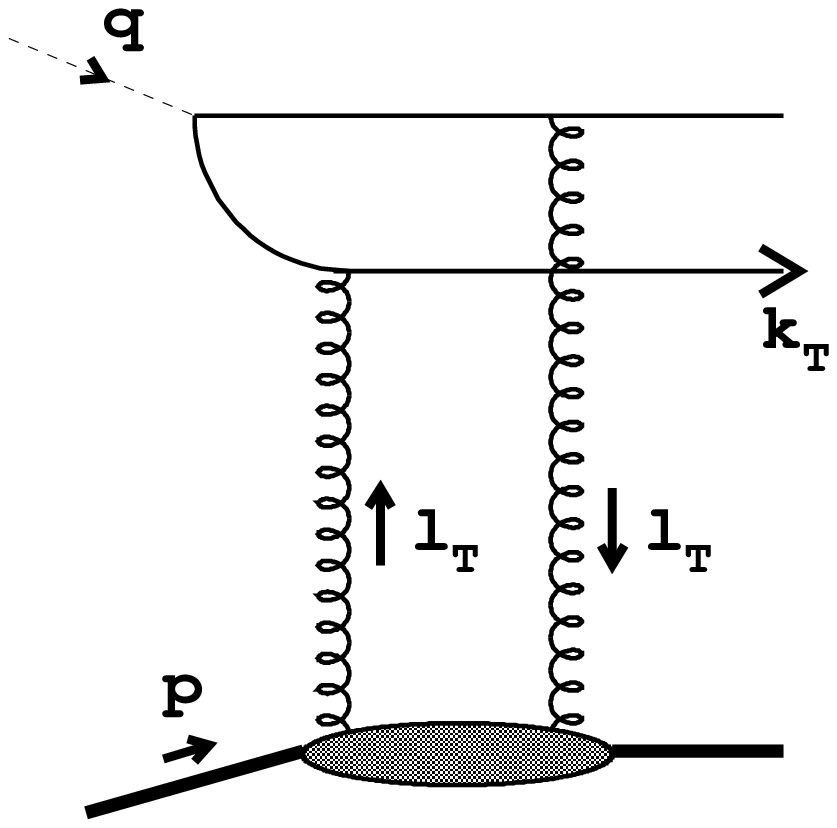,width=6cm} 
         \epsfig{figure=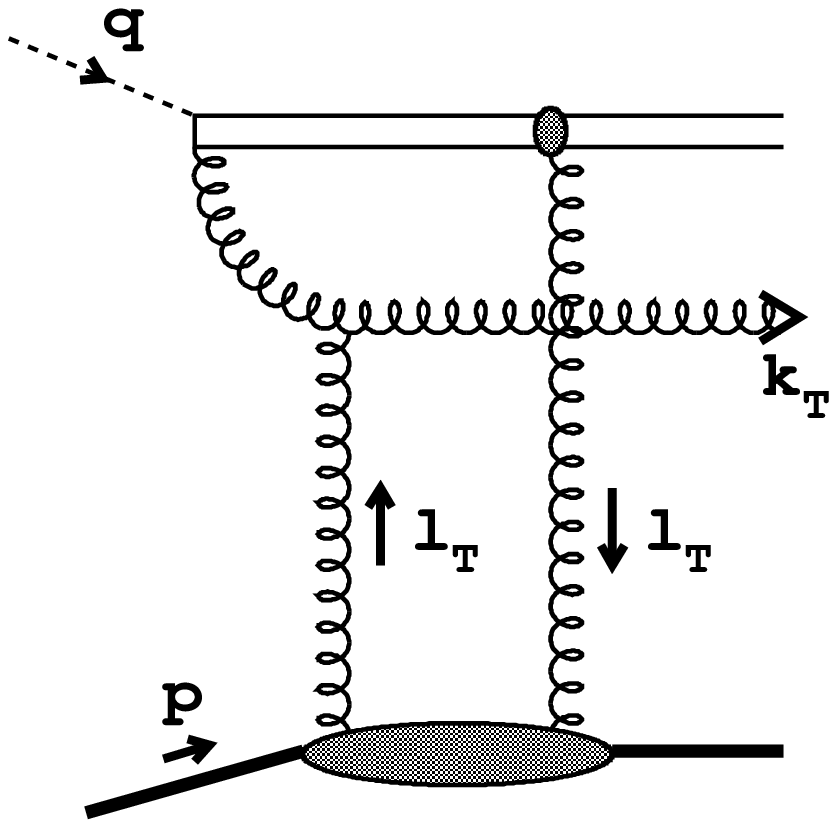,width=6cm}
           }
\vspace*{-0.5cm}
\caption{\it The diffractive $q\bar{q}$ and $q\bar{q}g$ contributions to
$F_2^{D(3)}$. 
\label{fig:model}}
\end{figure}

Following the idea of the analysis~\cite{BEKW}, 
the diffractive structure function $F_2^{D(3)}$ is the sum
of the three contributions shown in Fig.~\ref{fig:model}, the $q\bar{q}$ 
production from transverse and longitudinal photons, and
the $q\bar{q}g$ production,
\be
\label{eq:f2dtot}
F_2^{D(3)}(\beta,Q^2,\xp)\,=\,
F^T_{q\bar{q}}\,+\,F^L_{q\bar{q}}\,+\,F^T_{q\bar{q}g},
\ee
where $T$ and $L$ refer to the polarization of
the virtual photon. For the $q\bar{q}g$ contribution only the transverse
polarization is considered, since the longitudinal counterpart has no leading
logarithm in $Q^2$. In this approach, the diffractive $q\bar{q}$ and
$q\bar{q}g$   systems  interact with the proton
like in the two gluon exchange model with 
the coupling  to the proton described by the {dipole cross section} given by 
the  saturation model~\cite{GBW1}.

We find the diffractive parton distribution in such a model by extracting the 
leading twist part, i.e. that which depends logarithmically on $Q^2$,  from 
Eq.~(\ref{eq:f2dtot}). In this case only $F^T_{q\bar{q}}$ and 
$F^T_{q\bar{q}g}$ contribute since the longitudinal contribution 
$F^L_{q\bar{q}}$ is higher twist~\cite{BEKW} suppressed by the additional 
power of $1/Q^2$.

The computation of the diffractive structure function (\ref{eq:f2dtot})
was presented in~\cite{GBW2}. Here we quote only the
final results. The transverse $q\bar{q}$ part is given by 
\beeq
\label{eq:ftqq}
F_{q\bar{q}}^{T}
\,=\,
\frac{3}{64\pi^4 B_D\,\xp}\;\sum_q e_q^2 \;
\frac{\beta^2}{(1-\beta)^3}\;
\int_0^{\frac{Q^2(1-\beta)}{4\,\beta}} dk^2\;
\frac{\displaystyle 1-\frac{2\beta}{1-\beta}\frac{k^2}{Q^2}}
{\displaystyle \sqrt{1-\frac{4\beta}{1-\beta}\frac{k^2}{Q^2}}}\;\;
\phi_1^2,
\eeeq
where
\be
\label{eq:phi}
\phi_1=\phi_{1}(k,\beta,\xp)
\;=\;k^2\,
\int_0^\infty dr\, r\, K_{1}\!\left(\sqrt{\frac{\beta}{1-\beta}}kr\!\right)\,
J_{1}(kr)\;  \hat{\sigma}(\xp,r)
\ee
and $K_{1}$ and $J_{1}$ are the Bessel functions.
In the saturation model the dipole cross section
$\hat{\sigma}=\sigma_0\,(1-\exp(-r^2/4R_0^2(\xp))$ where $R_0^2\sim \xp^\lambda$. The
leading twist part is found by neglecting the factor
\be
\label{eq:factor}
\frac{\beta}{1-\beta}\,\frac{k^2}{Q^2}\,=\,z(1-z)\ll 1
\ee
under the integral in (\ref{eq:ftqq}) and taking the upper limit of the 
integration to infinity. Strictly speaking, energy
conservation is violated in this case, but the  corrections are higher
twists and  are neglected because of their smallness.
In Eq.~(\ref{eq:phi}) $z,\, (1-z)$ are the longitudinal momentum
fractions of the final state quarks with respect to the photon momentum.
Thus, the leading twist part of $F_{q\bar{q}}^{T}$ corresponds to the {\it 
aligned jet configurations} of the $q\bar{q}$ pair in the proton rest frame
since $z~\mbox{\rm or}~(1-z)\approx 0$. By the comparison of the leading twist 
part with Eq.~(\ref{eq:dpd2}) we find the {\it diffractive quark 
distribution}
\be
\label{eq:pdq}
q^D(\beta,\xp)\,=\,
\frac{3}{128\pi^4 B_D\,\xp}\;
\frac{\beta}{(1-\beta)^3}\;
\int_0^{\infty} dk^2\;
\phi_1^2(k,\beta,\xp),
\ee
where $B_D$ is the diffractive slope resulting from the $t$-integration of 
$F_2^{D(4)}$.

\begin{figure}[t]
   \vspace*{-1cm}
    \centerline{
     \epsfig{figure=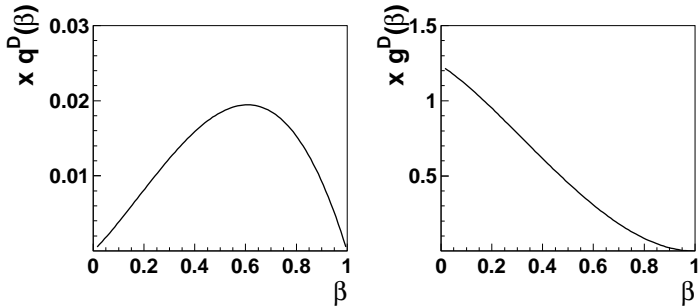,width=10cm}
               }
    \vspace*{-0.5cm}
\caption{Diffractive quark 
and gluon distributions (multiplied by $x=\beta\, \xp$) as a function of 
$\beta$ for $\xp=0.0042$ at an
initial scale $Q_0^2$. }
\label{fig:dpd}
\end{figure}

The $q\bar{q}g$ contribution was computed in~\cite{MARK2,GBW2} assuming strong
ordering in transverse momenta of the gluon and the $q\bar{q}$ pair,
i.e. $k_{\perp g} \ll k_{\perp q} \approx k_{\perp \bar{q}}$. This assumption
allows to treat the $q\bar{q}g$ system as a $gg$ dipole in the transverse
configuration space $r$. From the point of view of the leading logarithmic
contribution in $Q^2$ we find~\cite{GBW3}
\be
\label{eq:ftqqg}
F_{q\bar{q}g}^{T}
\,=\,
2\,
\sum_q e_q^2\;\beta\,
\frac{\alpha_s}{2\pi}\; \log\frac{Q^2}{Q_0^2}\; 
\int_\beta^1 \frac{d\ovbeta}{\ovbeta}\;\frac{1}{2}
\left[\left(1-\frac{\beta}{\ovbeta}\right)^2
+\left(\frac{\beta}{\ovbeta}\right)^2\right]\,
g^D({\beta},\xp)
\ee
which formula allows to define the {\it diffractive gluon distribution} 
\be
\label{eq:pdg}
g^D(\beta,\xp)
\;=\;
\frac{81}{256\pi^4 B_D\,\xp}\;
\frac{\beta}{(1-\beta)^3}\;
\int_0^{\infty} dk^2\; \phi_2^2(k,\beta,\xp),
\ee
where $\phi_2$ is given by Eq.~(\ref{eq:factor}) with the subscript 1 replaced 
by 2. Notice the lack of the $Q^2$-dependence in both (\ref{eq:pdq}) and 
(\ref{eq:pdg}).  This may be viewed as  a consequence of
not having included ultraviolet divergent corrections which would require a
cutoff. With those corrections the parton distributions become
$Q^2$-dependent and evolution would relate the distributions at different
$Q^2$. Still, we may use the found DPD
as the input for the DGLAP evolution equations
at some initial  scale $Q_0^2$. 

\section{Regge factorization}

The scaling property of the dipole cross section in the saturation 
model~\cite{GBW1},
i.e. that $\hat{\sigma}$ is a function of the ratio $r/R_0(\xp)$,  leads to
the factorized  $\xp$-dependence of the DPD, similar to Regge factorization.
Introducing $\hat{k}=k R_0(\xp)$ and $\hat{r}=r/R_0(\xp)$ in (\ref{eq:pdq}) and
(\ref{eq:pdg}) and assuming $Q_0^2$ fixed, we find 
\beeq
q^D(\beta,\xp) &=& \frac{1}{\xp R_0^2(\xp)}\;f_{q}(\beta)
\\ \nonumber
\\
g^D(\beta,\xp) &=& \frac{1}{\xp R_0^2(\xp)}\;f_{g}(\beta).
\eeeq
The DGLAP evolution does not affect the
$\xp$-dependence and  the factorized form is valid for any scale 
$Q^2$. Thus, the leading twist part of $F_2^{D(3)}$ 
\be
\label{eq:poms1}
F_2^{D(3)(LT)}\,\sim\,
\xp^{-1-\lambda},
\ee
where the parameter $\lambda=0.29$ of  the saturation model was  found in the 
analysis of inclusive $F_2$. This value interpreted in terms of 
the $t$-averaged  pomeron intercept,
$F_{2}^{D(3)(LT)}\sim \xp^{1-2\,\overline{\alpha_\funp}}$,  gives
$\overline{\alpha_\funp}=1.15$ which is in remarkable agreement with the value 
$1.17$ found by H1 \cite{H197} and $1.13$ by ZEUS \cite{ZEUS99}.

Thus,  the saturation model gives effectively the result which coincides with 
the Regge approach to DIS diffraction, although the physics behind is 
completely different. The relative hardness of the intrinsic scale 
$1/R_0(\xp)\sim 1~\mbox{\rm GeV}$  
in the saturation model suggests that DIS diffraction is a semihard process 
rather than a soft process as Regge theory would require.

\section{Comparison with the data}

We use the following model for the diffractive structure function
\be
\label{eq:newanal}
F_2^{D(3)}\,=\,F_2^{D(3)(LT)}\,+\,F_{q\bar{q}}^{L},
\ee
where $F_2^{D(3)(LT)}$ is given by Eq.~(\ref{eq:dpd2}) 
with the DGLAP evolution from the staring conditions given by
Eqs.~(\ref{eq:pdq}) and (\ref{eq:pdg}), and twist-4
$F_{q\bar{q}}^{L}$ is defined as in the model (\ref{eq:f2dtot}), 
see~\cite{GBW3} for an explicit formula. The latter contribution
is crucial in the region  $\beta\approx 1$. The starting point
$Q_0^2\approx 1.5~\mbox{\rm GeV}^2$ for the evolution was found.
to get a good description of the data. The leading
$\log(Q^2)$ evolution with $N_f=3$ flavours and $\Lambda=200~\mbox{\rm MeV}$ 
was assumed.

\begin{figure}[t]
   \vspace*{-1cm}
    \centerline{
     \epsfig{figure=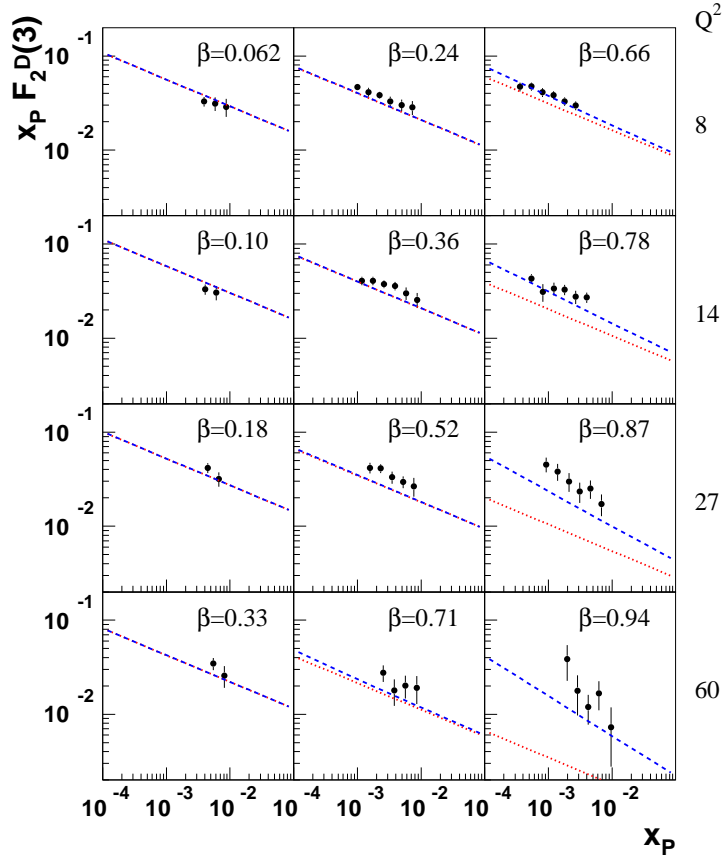,width=10cm}
               }
    \vspace*{-0.5cm}
\caption{
The comparison with ZEUS data. 
The dashed lines correspond to the model (\ref{eq:newanal}). 
The leading twist contribution is shown by the dotted lines.
}
\label{fig2}
\end{figure}

The results of the comparison with the ZEUS data are presented in 
Fig.~\ref{fig2}. A similar comparison with the H1 data is presented 
in~\cite{GBW3}.
Notice a reasonable good agreement. The description in the region of large
$\beta$ and $Q^2$, however, exhibits some problems which might be
removed if such effects like twist-4 evolution or/and skewedness of the gluon 
distribution is taken into account.

\section*{Acknowledgments}
This research has been supported in part by the Polish KBN
grant No.  5 P03B 144 20. The financial support of   Deutsche 
Forschungsgemeinschaft is gratefully acknowledged.


\begin{thebibliography}{99}

\bibitem{COL98} J. C. Collins,
         {\it Phys. Rev.} {\bf D57} (1998) 3051,  
          Erratum-{\it ibid.} {\bf D61} (2000) 019902.

\bibitem{H197} H1 Collaboration, C. Adloff {\it et al.},
               {\it Z. Phys.} {\bf C76} (1997) 613.

\bibitem{ZEUS99} ZEUS Collaboration, M. Derrick {\it et al.},
               {\it Eur. Phys. J.} {\bf C6} (1999) 43.

\bibitem{REV} M. W\"usthoff and A.D. Martin,
                 {\it J. Phys.} {\bf G25} (1999) R309;\\
                  A. Hebecker, 
                 {\it Phys. Rep.} {\bf 331} (2000) 1.

\bibitem{GBW1} K. Golec--Biernat and  M. W\"usthoff,
            {\it Phys. Rev.} {\bf D59} (1999) 014017.

\bibitem{BEKW}  J. Bartels, J. Ellis, H. Kowalski and M. W\"usthoff,
                  {\it Eur. Phys. J.} {\bf C7} (1999) 443.

\bibitem{GBW2} K. Golec--Biernat and  M. W\"usthoff,
           {\it Phys. Rev.} {\bf D60} (1999) 114023.


\bibitem{MARK2} M. W\"usthoff, {\it Phys. Rev.} {\bf D56} (1997) 4311.


\bibitem{GBW3} K. Golec--Biernat and  M. W\"usthoff, 
           {\tt hep-ph/0102093}. 
\end{thebibliography}
\end{document}